\newcommand{\taupg}{\mbox{ e$^{+}$e$^{-}\rightarrow\tau^{+}\tau^{-}(\gamma)$}}
\def\mepem{\mathrm{e^+e^-}}

\def\epem{$e^+e^-$}

\def\mmpmm{\mu^+\mu^-}

\def\tlnn{\ensuremath{\tau\rightarrow\ell\bar{\nu}_{\ell}\nu_{\tau}}}
\def\gwt{\ensuremath{g_{\tau}}}
\def\gwm{\ensuremath{g_{\mu}}}
\def\gwe{\ensuremath{g_{\e}}}
\def\gwl{\ensuremath{g_{\ell}}}
\def\mtau{\ensuremath{m_{\tau}}}
\def\mw{\mbox{$m_\mathrm{W}$}}
\def\mz{\mbox{$m_\mathrm{Z}$}}
\def\menn{\ensuremath{\mu\rightarrow\e\bar{\nu}_{\e}\nu_{\mu}}}
\def\als{\ensuremath{\alpha_s}}

\def\dnpqcd{\ensuremath{\delta_{NP}}}
\def\sew{\ensuremath{S_{EW}}}

\def\tlnn{\ensuremath{\tau\rightarrow\ell\bar{\nu}_{\ell}\nu_{\tau}}}
\def\tenn{\ensuremath{\tau\rightarrow\e\bar{\nu}_{\e}\nu_{\tau}}}
\def\tmnn{\ensuremath{\tau\rightarrow\mu\bar{\nu}_{\mu}\nu_{\tau}}}
\def\brlep{\ensuremath{{\cal{B}}(\tau\rightarrow\ell\bar{\nu}_{\ell}\nu_{\tau})}}
\def\brmu {\ensuremath{{\cal{B}}(\tmnn)}}
\def\brel {\ensuremath{{\cal{B}}(\tenn)}}
\def\rtau {\ensuremath{R_{\tau}}}
\def\mmu  {\ensuremath{m_{\mu}}}
\def\mtau {\ensuremath{m_{\tau}}}
\def\ttau {\ensuremath{\tau_{\tau}}}
\def\tmu  {\ensuremath{\tau_{\mu}}}
\def\Vud{\ensuremath{\vert V_{\mathrm{ud}} \vert}}
\def\Vus{\ensuremath{\vert V_{\mathrm{us}} \vert}}
\documentclass[12pt,a4paper,dvips]{article}
\usepackage{a4p}
\usepackage{cite,mcite}
\usepackage{graphicx}
\usepackage{physics }
\usepackage{l3_title,ifthen,Lep}
\journalname{Phys. Lett. B}
\date{\today}
\preprint{2000-XXX}
\Lep{1}
\newlength{\capindent}
\newlength{\capwidth}
\newlength{\figwidth}
\newcommand{\icaption}[2][!*!,!]{\hspace*{\capindent}%
  \begin{minipage}{\capwidth}
    \ifthenelse{\equal{#1}{!*!,!}}%
      {\caption{#2}}%
      {\caption[#1]{#2}}
  \end{minipage}}

\def\antibar#1{#1\bar{#1}}
\def\qqbar{\antibar{\mathrm{q}}}
\def\e{\mathrm{e}}
\def\sint{\sigma_{\mbox{\scriptsize int}}}
\def\sip{\sigma_{\mbox{\scriptsize ip}}}
\def\sxb{\sigma_{x}}
\def\syb{\sigma_{y}}
\def\sms{\sigma_{\mbox{\scriptsize ms}}(p)}
\def\smssq{\sigma_{\mbox{\scriptsize ms}}^2(p)}
\def\micron{\mu \mbox{m}}
\def\mepem{\mathrm{e^+e^-}}

\def\epem{e^+e^-}

\def\mmpmm{\mu^+\mu^-}

\def\tlife{\tau_{\tau}}
\def\fs   {\mbox{fs}}
\begin{document}
\begin{titlepage}
  \title{Measurement of the Lifetime of the Tau Lepton}
  \author{The L3 Collaboration}
  \begin{abstract}
    The tau lepton lifetime is measured with the L3 detector 
    at LEP using the complete data 
    taken
    at centre-of-mass energies
    around the Z pole
    resulting in
    $\tlife\,=\,293.2\,\pm\,2.0~\mbox{(stat)}\,\pm\,1.5~\mbox{(syst)}~\fs$.
    The comparison of this result      
    with the muon lifetime supports 
    lepton universality of the weak charged current
    at the level of six per mille.
    Assuming lepton universality, the value of the strong coupling 
    constant, $\alpha_s$ 
    is found to be $\alpha_s(\mtau^2)$ = 0.319 $\pm$ 0.015~\mbox{(exp)}
    $\pm$ 0.014~\mbox{(theory)}.
\end{abstract}
\submitted
\end{titlepage}  

\section {Introduction}

In the Standard Electroweak Model~\cite{GWS1,*GWS2,*GWS3},
the couplings of the leptonic charged and neutral currents to the gauge bosons
are 
independent of the lepton generation. 
Measurements of
the lifetime, $\ttau$,
and the leptonic branching fractions, $\brlep$,
of the tau lepton
provide a test of this universality hypothesis
for the charged current.
The leptonic width 
of the tau lepton~\cite{Sirlin73,*Marciano88},
\begin{eqnarray}
\Gamma(\tlnn) & \equiv & \frac{\brlep}{\ttau} \\
              & = & \frac{\gwt^2\gwl^2}{\mw^4}\frac{\mtau^5}{96 (4\pi)^3}
                    (1+\epsilon_P)\,(1+\epsilon_{rad})\,(1+\epsilon_{q^2}), 
\end{eqnarray}
where $\ell=$e,$\mu$,
depends on the coupling constants of the tau lepton
and the lighter lepton to the
W boson, $\gwt$ and $\gwl$, respectively.
Here
$\mtau$ and $\mw$ are the masses
of the tau lepton  and the W boson.
The quantities $\epsilon_P$, $\epsilon_{rad}$ and $\epsilon_{q^2}$
are small corrections resulting from phase-space, radiation
and the W propagator, respectively. 
For the decay of the muon, $\menn$, the same formula
applies, with the muon mass and coupling, $\mmu$ and $\gwm$,
replacing those of the tau.
The comparison of the tau lifetime and leptonic branching fractions
with the muon lifetime 
gives a direct measurement of 
the ratios $\gwt/\gwe$ and $\gwt/\gwm$. 
 
Tau decays into hadrons are sensitive
to the strong coupling constant, $\als$, at the tau mass scale. 
Assuming universal coupling constants of
the different lepton species to the
W boson, the ratio of the 
hadronic width to the leptonic  width, $\rtau$, can be expressed as:
\begin{equation}
\rtau = \left(\frac{\mmu}{\mtau}\right)^5 \frac{\tmu}{\ttau} - C,
\end{equation}
where $C=1.9726$ contains the same corrections mentioned above.

$\rtau$ is calculated in perturbative QCD
~\cite{Gorishny91,Braaten92,Kataev95}:
\begin{equation}
  \label{eq:als}
  \rtau = 3 (\Vud^2 +\Vus^2)  \sew (1 + \frac{\als}{\pi}\,+\,
          5.2023 (\frac{\als}{\pi})^2 \,+\,
          26.366 (\frac{\als}{\pi})^3 \,+\,
          (78+d_3)(\frac{\als}{\pi})^4\,+\, \dnpqcd ).
\end{equation}

\noindent
The quantities $\Vud$ and $\Vus$ are elements of the Cabibbo-Kobayashi-Maskawa
(CKM)
quark mixing matrix~\cite{ckm1,*ckm2}, 
$\sew$~\cite{Sirlin82} and $\dnpqcd$~\cite{Braaten92,neubert96} describe
electroweak radiative corrections
and 
non-perturbative QCD contributions, respectively. The quantity
$d_3$ is 
estimated as $d_3$ = 27.5~\cite{Kataev95}.  
The perturbative theoretical uncertainty is taken to be
$d_3$ = 27.5$\pm$27.5. 

This paper presents a measurement of the tau lepton lifetime with 
the L3 detector at LEP, using data taken in 1995 at the $\Zo$ pole.
Furthermore, data from 1994 are re-analysed using an improved
calibration and alignment of the central tracker. 
The lifetime is measured 
from the decay length in three-prong tau decays and
from the impact parameter of one-prong tau decays\footnote{An $N$-prong 
tau lepton decay indicates a decay with $N$ charged particles in the final 
state.}.
The results are combined with our previously published analyses
on data collected from 1990 to 1993~\cite{l3_1993_3,*mblife1}.
Previous
measurements of the tau lepton lifetime have been reported 
by other experiments
\cite{others}.

\section{L3 Detector}

The L3 detector is described in Ref.~\cite{l3_1990_1}.
This measurement is based primarily
on the information obtained from the central 
tracking system, which is composed of a 
Silicon Microvertex Detector (SMD)~\cite{l3_1994_10,*smd_09},
a Time Expansion Chamber (TEC) and a Z-chamber.  
The SMD is made of two concentric
layers of double-sided silicon detectors, placed at about 6 and 8~cm from 
the beam line.
Each layer provides a two-dimensional position measurement, with a resolution
of $7~\mbox{and}~14~\mu \mbox{m}$ for normally
incident
tracks, in the directions perpendicular
and parallel to the beam direction, denoted as
$(x,y)$ and $z$ coordinates,
respectively. 
The TEC consists of two coaxial cylindrical
drift chambers with 12~inner and 24~outer sectors. The sensitive region is
between 10 and 45~cm in the radial direction, with 62 anode wires having
a spatial resolution of approximately $50~\mu \mbox{m}$ in the plane 
perpendicular to the beam axis. The Z-chamber, which is
situated just outside the TEC,
provides a coordinate measurement along the beam axis direction. 

\section{Event Sample}

For this measurement data collected in 1994 and 1995 are used, which correspond
to an integrated luminosity of 49 pb$^{-1}$ and 31 pb$^{-1}$, respectively.

For  efficiency  and  background   estimates,  Monte  Carlo  events  are
generated  using  the  programs  KORALZ~\cite{koralz1}  for $\rm e^+ e^-
\rightarrow       \mu^+      \mu^-       (\gamma)$      and      \taupg,
BHAGENE~\cite{bhagene1,*bhagene2}  for $\rm e^+ e^- \rightarrow  e^+ e^-
(\gamma)$,  DIAG36~\cite{diag36}  for $\rm e^+ e^- \rightarrow e^+ e^- f
\bar{f}  $, where  $\rm f  \bar{f} $ is $\rm e^+  e^-$,  $\mu^+  \mu^-$,
$\tau^+  \tau^-$ or $\rm q  \bar{q}$,  and  JETSET~\cite{my_jetset}  for
$\ee\ra\qqbar(\gamma)$.  The Monte  Carlo  events are  passed  through a
full  detector  simulation  based on the GEANT  program~\cite{my_geant},
which  takes  into  account  the  effects  of  energy   loss,   multiple
scattering, showering and small
time dependent detector inefficiencies. These
events are  reconstructed  with the
same program used for the data.
The  number of Monte  Carlo  events in each  process  is about ten times
larger than the corresponding data sample.
  
Tau lepton pairs originating from $\Zo$ 
decays are
characterised by two low multiplicity, highly collimated jets in
the detector.
The selection of $\taupg$ events is described in detail
in Ref.~\cite{life94}; here only a general outline
is given.
In order to have high-quality reconstruction of the tracks, 
events are accepted within a fiducial 
volume defined by $|\cos \theta_{t}| < 0.72$,
where the polar angle $\theta_{t}$ is given by the thrust axis 
of the event with respect to the electron beam 
direction. The events must have at least two jets, and
the number of tracks in each jet
must be less than four.
The background from $\mepem\rightarrow\mepem~(\gamma)$
events is reduced by
requiring the total energy deposited in the electromagnetic calorimeter 
to be less than $75\%$
of the centre-of-mass energy.
To reduce
background from
$\mepem\rightarrow\mmpmm~(\gamma)$
the
sum of the absolute momenta
measured in the muon spectrometer must be less than  
$70\%$ of the centre-of-mass energy.
If muons are not reconstructed in the muon chambers they
are identified by an energy deposit in the
calorimeters
which is characteristic of a minimum ionising particle.
If this is the case for
one jet,
the opposite jet is required to exhibit a 
hadronic signature.
This rejects dimuon as well as cosmic-ray events.
The cosmic-ray background is further reduced by requiring 
a scintillation counter hit within 5~ns of the beam crossing.
In addition, the distance of closest approach to the 
interaction point measured
by the muon chambers must be less than two standard deviations
of the resolution.

Following this procedure, $29679$ and $13294$ events are selected 
from the data collected in 1994 and 1995, respectively. The selection 
efficiency in the fiducial volume is estimated to be 76\%. The purity
of the tau pair sample is 98\%.

\section{Tracking}

\begin{figure}
    \includegraphics[width=0.49\textwidth]{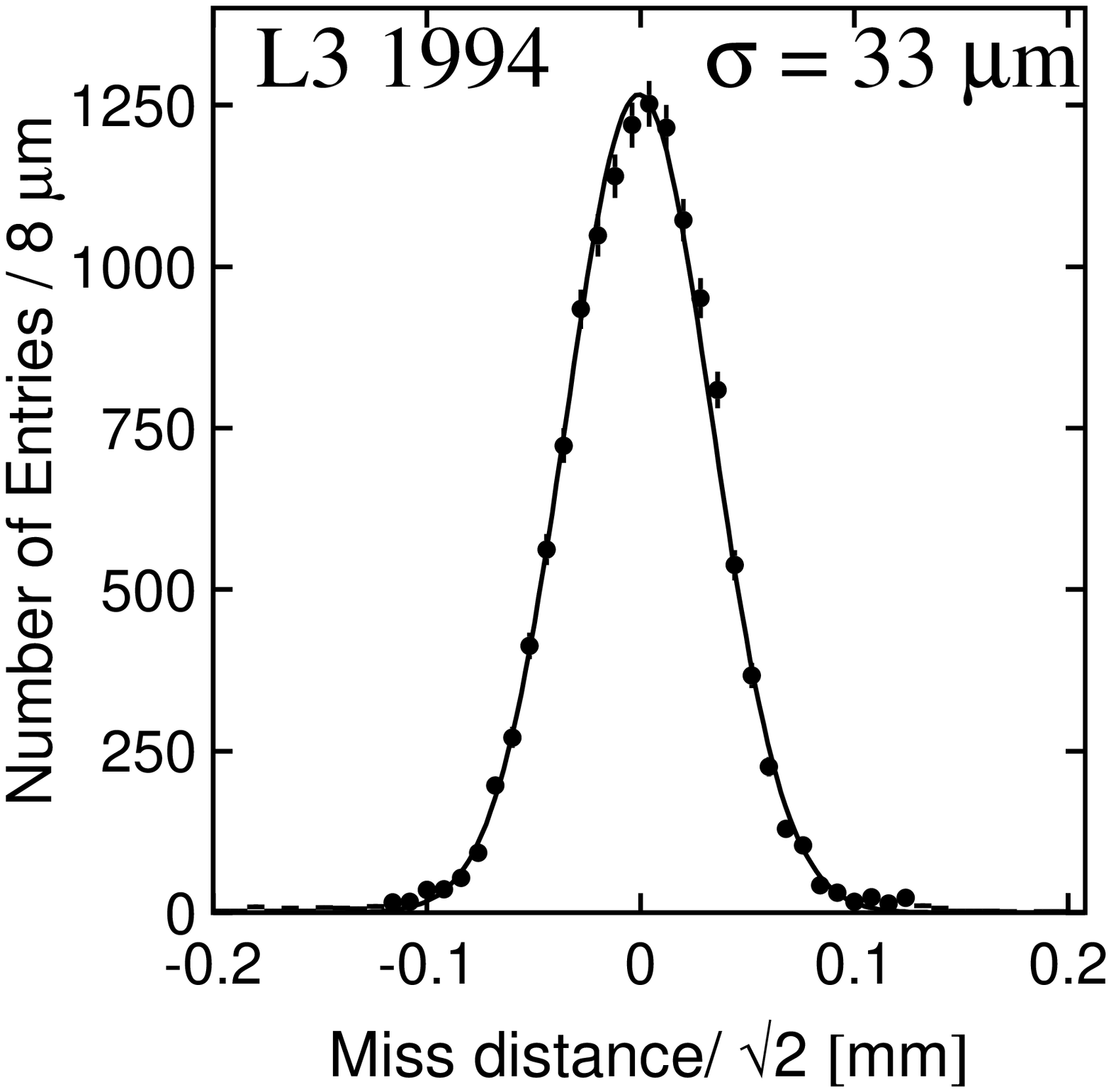}
    \includegraphics[width=0.49\textwidth]{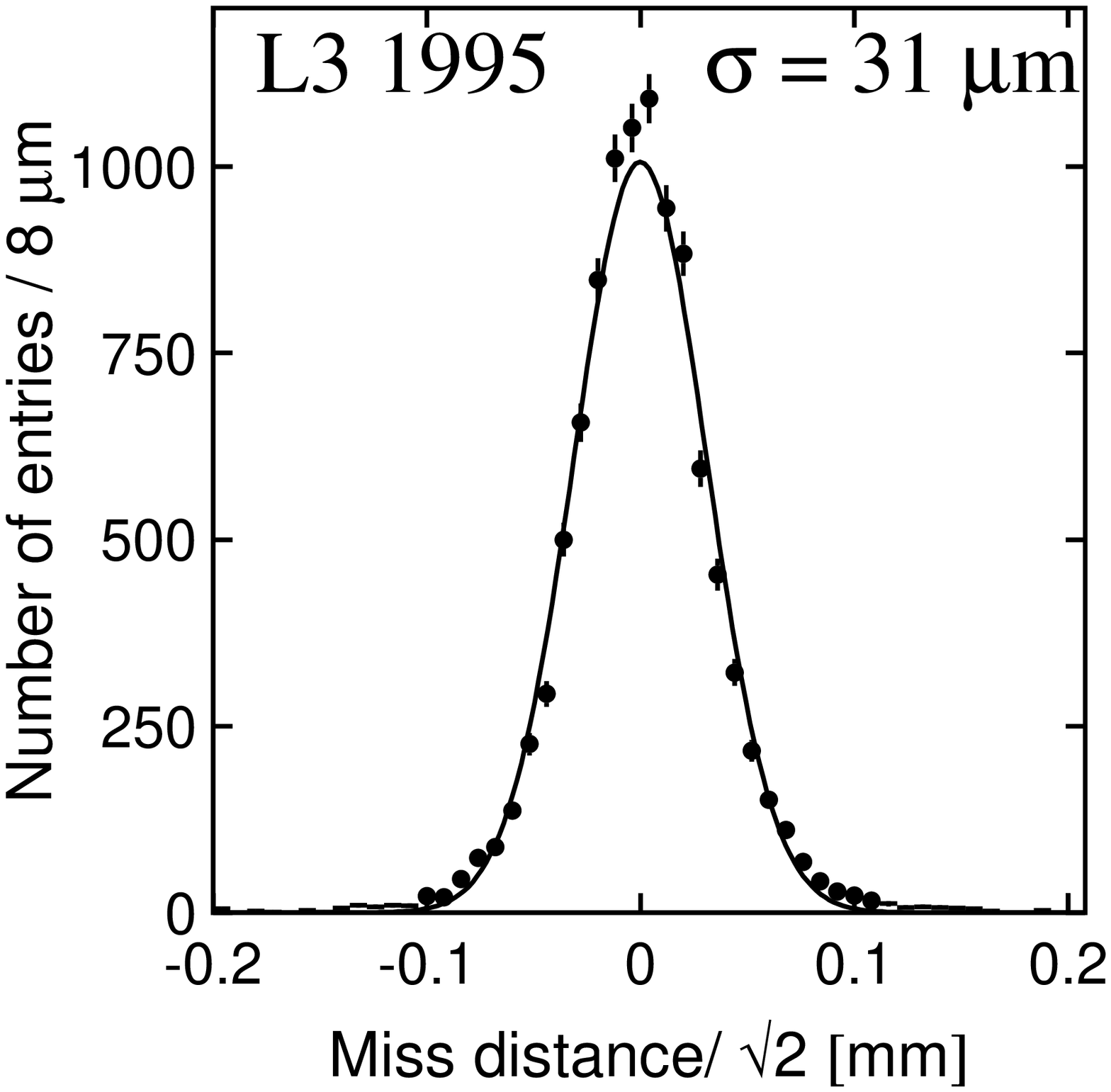}
    \includegraphics[width=0.49\textwidth]{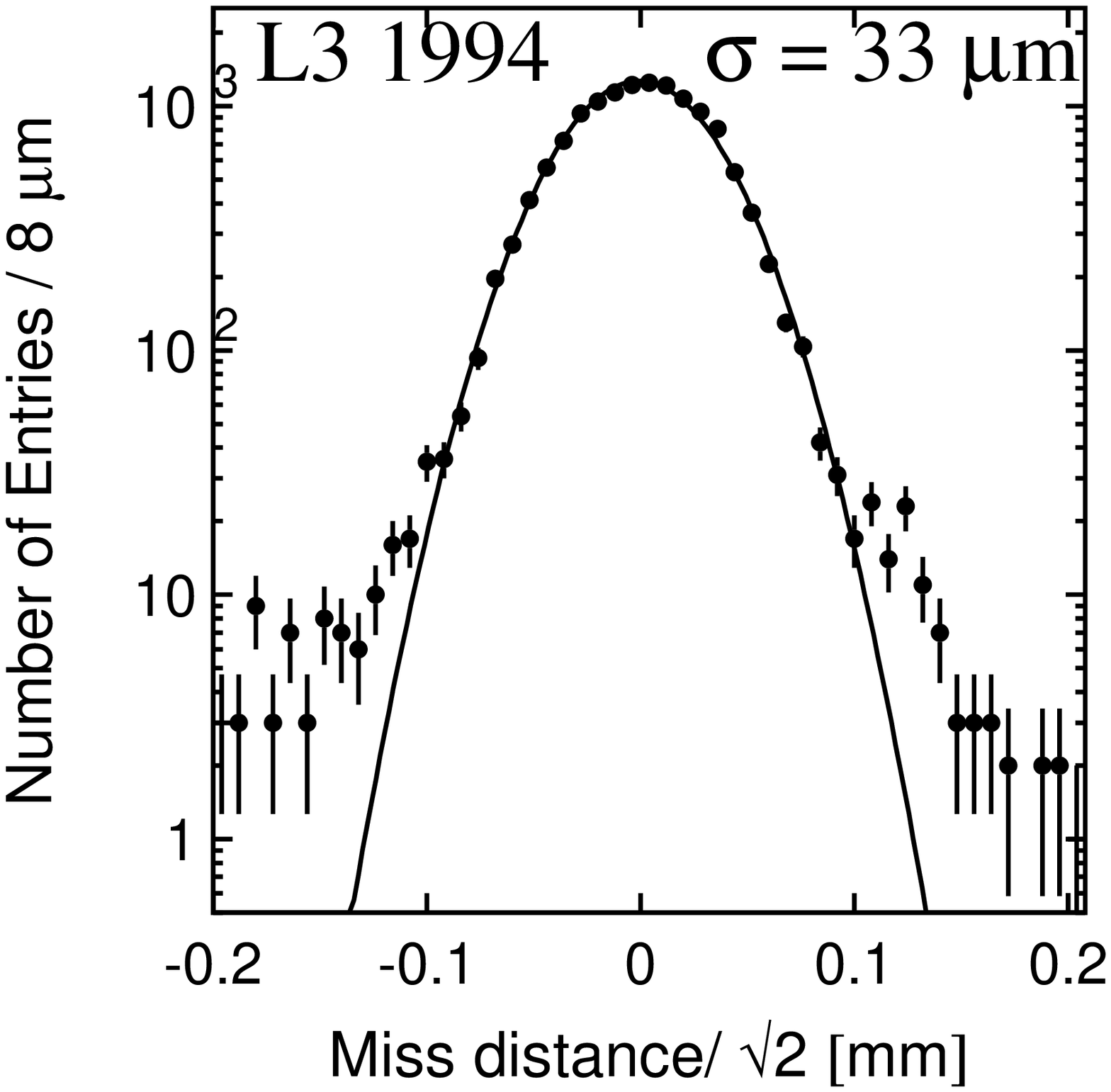}
    \includegraphics[width=0.49\textwidth]{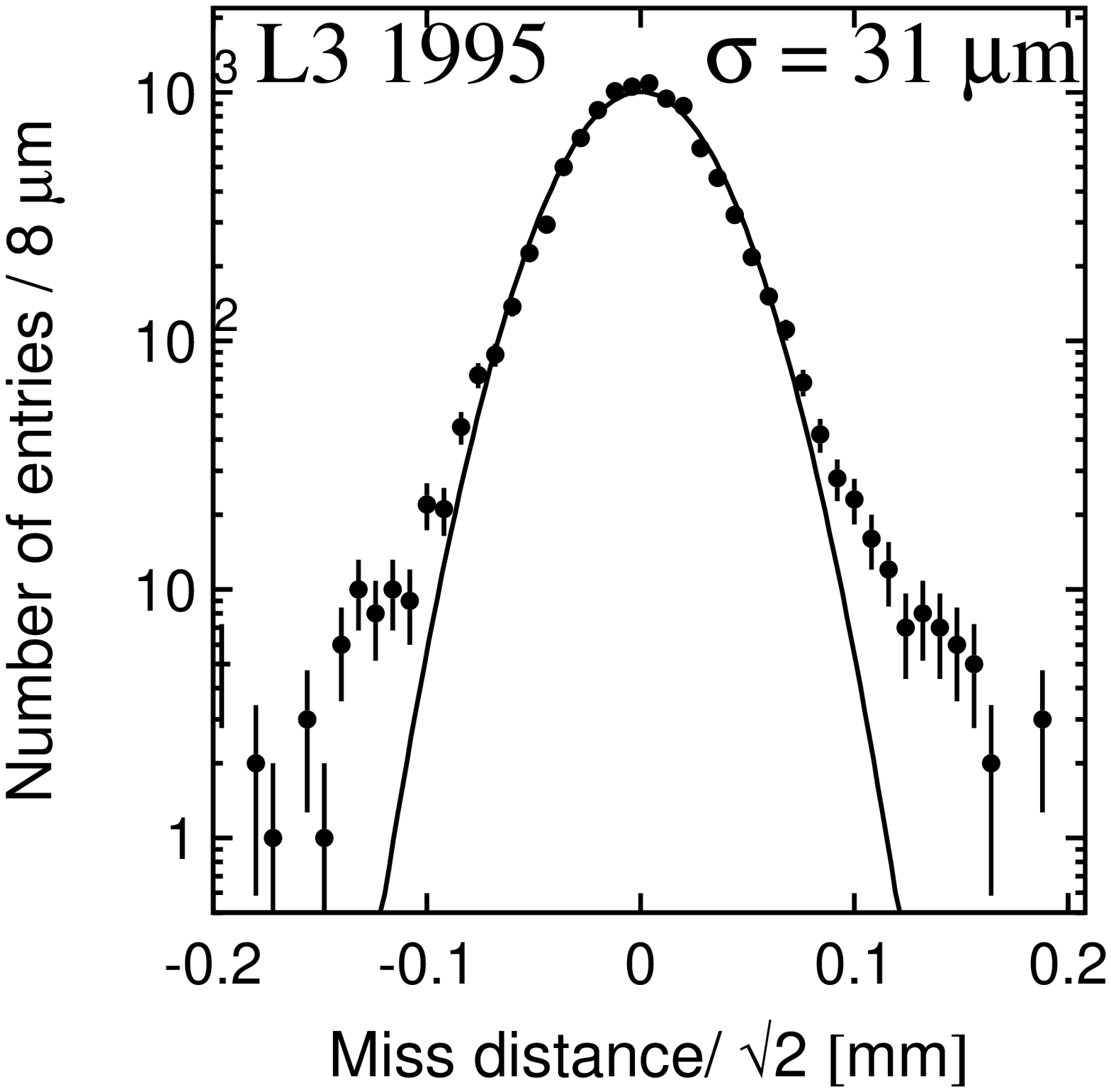}
    \caption[]{\label{fig:reso}
             Miss distance distributions from 1994~(left) and 
             1995~(right); Bhabha and dimuon events
             are shown in linear scale (upper)
             and logarithmic scale (lower). Dots are data
             and the solid line is the result of a fit with 
             a Gaussian. 
            }
\end{figure}

For this measurement a high quality 
of the track reconstruction is essential. A prerequisite 
is the control
of the alignment 
between SMD and TEC and the drift time to drift distance
calibration for the TEC.
These calibrations, alignments
and the estimation of resolution functions are  
performed with a clean sample of Bhabha and dimuon events,
where tracks are known to originate
from a common vertex.
Particular effort is invested
in the individual alignment of each sensor of the SMD
and in the calibration of the boundaries of the TEC sectors.
The procedure is described in detail in Ref.~\cite{colijn.ph.d}.
The performance of the track reconstruction
is estimated from the distance
between the two tracks
at the vertex projected into
the $(x,y)$ plane.
This quantity,
called miss distance, 
is independent
of the size and position of the $\e^+ \e^-$ interaction region.
The distributions of miss distance for Bhabha and 
dimuon events collected during 1994 and 1995 are shown in 
Figure~\ref{fig:reso}. A Gaussian
function is fitted to both distributions, from which an intrinsic
resolution 
$\sint = 33~\micron$ and $31~\micron$ is estimated
for 1994 and 1995, respectively.

To guarantee good tracks for the analysis, the following cuts are
made:
\begin{itemize}
\item{Number of hits in the TEC $\geq 30$. This ensures a good curvature 
    measurement.
    }
\item{Number of SMD hits in the $(x,y)$ plane $\geq 1$.
      This criterion selects tracks
      for which the error in the extrapolation
      to the vertex is well described by $\sint$.
    }
\item{Transverse momentum, $|p_{t}|\geq 500~\mbox{MeV}$. 
    Tracks with lower momenta have a larger
    uncertainty in the extrapolation
    to the vertex due to multiple Coulomb scattering in the SMD.    
    }
\item{Probability, $P(\chi^2)$, of the track fit larger than 1\%.
      This requirement rejects bad fits.
    }
\end{itemize}

\section{Decay Length Method}

\begin{figure}[htb]
   \begin{center}
    \includegraphics[width=10.5cm]{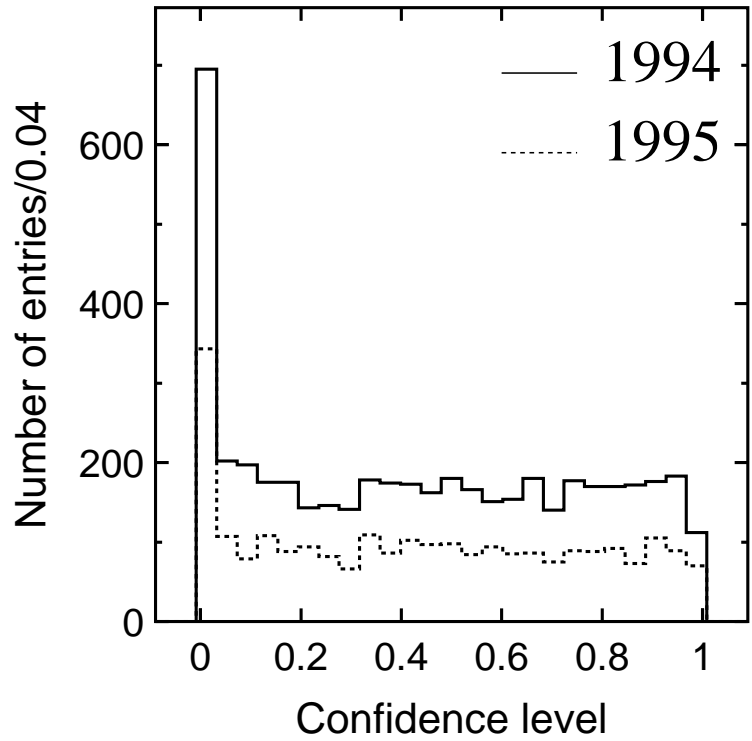}
   \end{center}
    \caption[]{\label{fig:pc2} 
             Confidence level, $P(\chi^2)$, of the secondary vertex 
             reconstruction. 
                              }
\end{figure}

For three-prong tau decays, the decay vertex of the tau is reconstructed and its
distance to the centre of the interaction region is
measured. The decay vertex is found from a minimisation with
respect to the vertex coordinates $(x_v,\,y_v)$ of the following $\chi^2$:
\begin{equation}
\chi^2\,=\,\sum_{i=1}^{N_{\mbox{\scriptsize track}}} \left(
  \frac{\delta_i(x_v,\,y_v)}{\sigma_{\delta_i}}
  \right)^2.
\end{equation}
In this equation,
 $\delta_i$ is the distance of closest approach of a
track to the decay vertex coordinates.
The error, $\sigma_{\delta_i}$, 
is the quadratic sum of the intrinsic
resolution, $\sint$,
and the error due to multiple Coulomb scattering, $\sms$. 
Figure~\ref{fig:pc2} shows the distribution of the
confidence level, $P(\chi^2)$,
of the fit.
Decay vertices with $P(\chi^2) < 1\%$ 
contain in most cases tracks with wrongly matched SMD hits
and  are rejected.
The remaining distribution is flat, indicating a good description of the 
errors.

\begin{figure*}[ht]
    \includegraphics[width=0.49\textwidth]{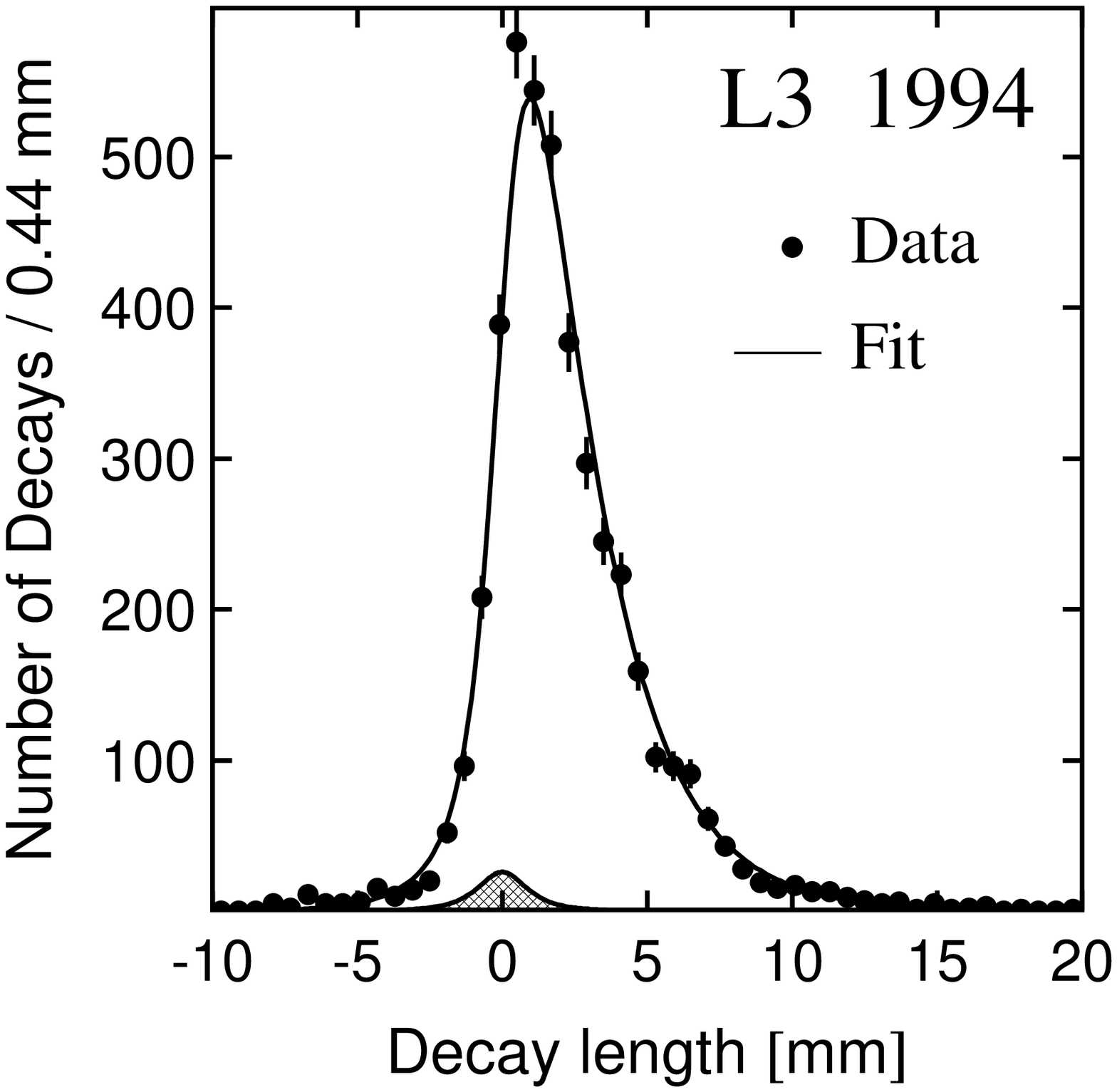}
    \includegraphics[width=0.49\textwidth]{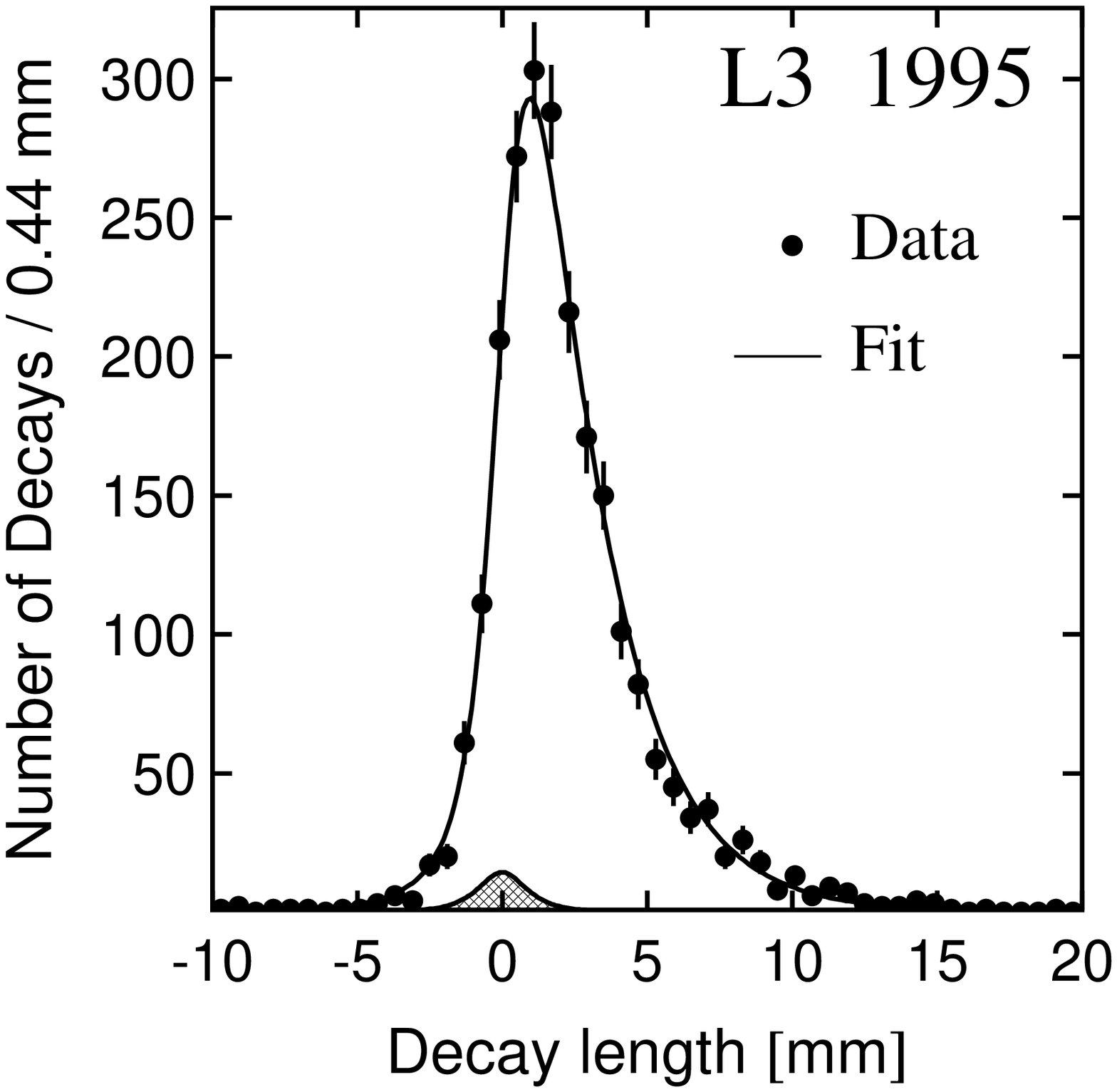}
    \includegraphics[width=0.49\textwidth]{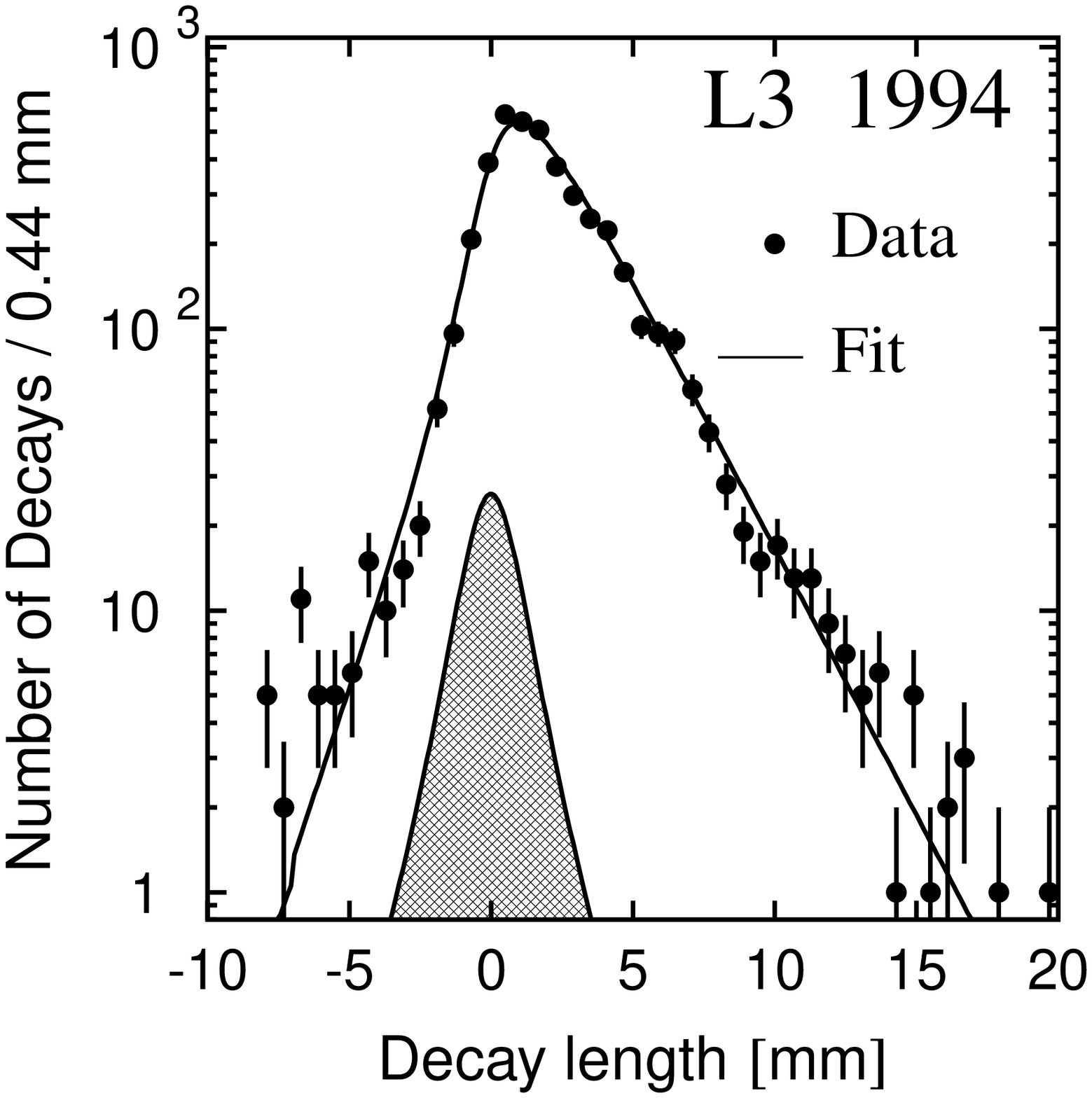}
    \includegraphics[width=0.49\textwidth]{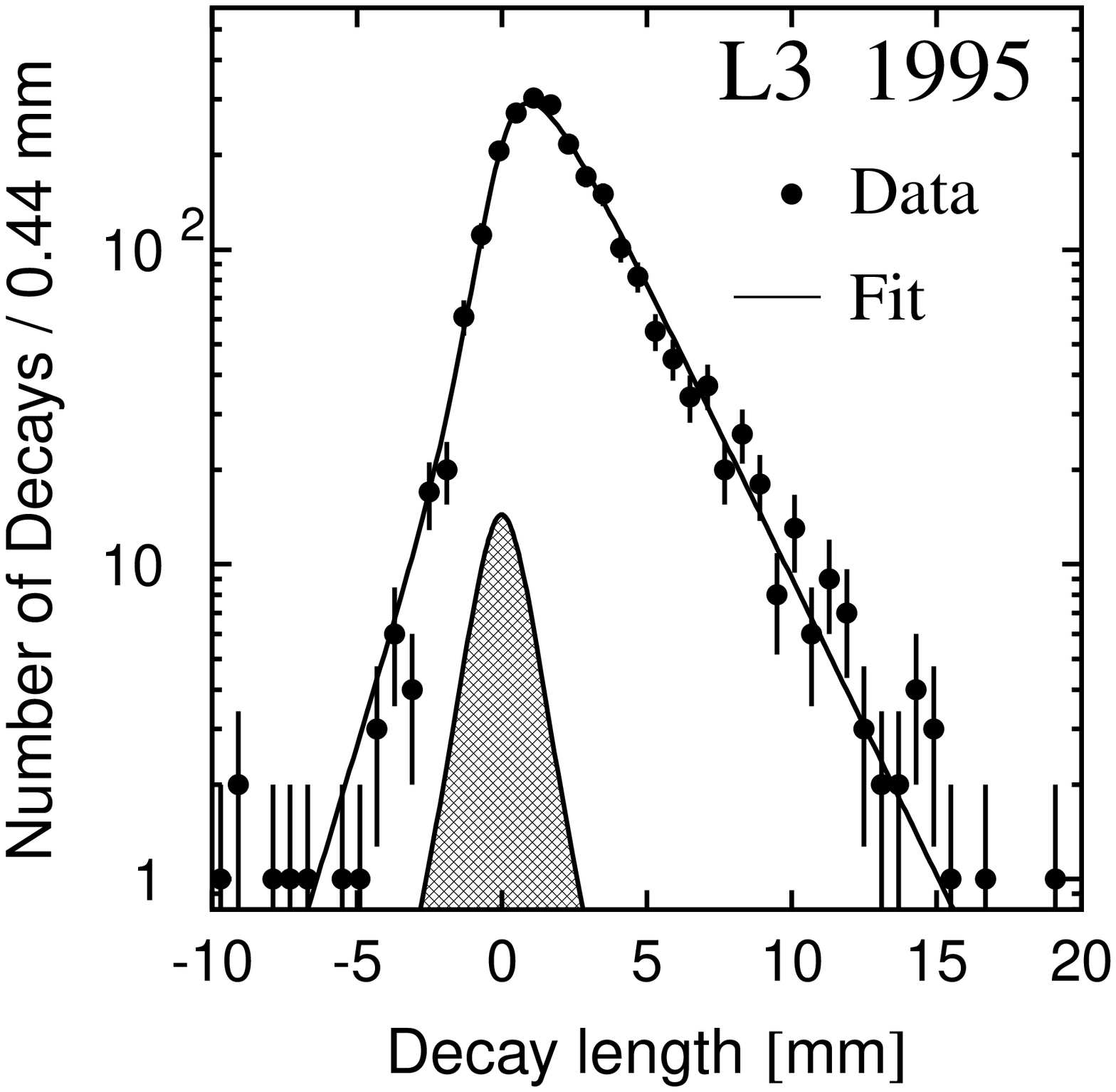}
    \caption[]{\label{fig:dl}
             Decay length distributions from 1994~(left) and
             1995~(right); Three-prong tau decays 
             are shown in linear scale (upper)
             and logarithmic scale (lower).
             The hatched areas represent the distributions
             of background events carrying no lifetime information.
        }
\end{figure*}

Figure~\ref{fig:dl} shows the decay length distribution for the data 
collected in 1994 and 1995. Only decay length values
in the range $[-10,\,20]~\mbox{mm}$ with an error better than 5~mm
are accepted for the lifetime determination.
The number of selected decays is $4306$ and 
$2314$ for 1994 and 1995 data, respectively.

To obtain the average decay length,  
an unbinned maximum likelihood fit is
applied to the observed decay
length distribution.
The likelihood function is calculated from a convolution 
of an exponential $E$,
describing the tau decay time using the average decay length
$\langle L \rangle$ as a parameter,
with a Gaussian resolution 
function $R$.
The likelihood function is written as:
\begin{equation}
{\cal{L}}\,=\, \prod_{i=1}^{N_{3p}} (1-f_B)\cdot E \otimes R + f_B \cdot B.
\end{equation} 
In this equation the product runs over the accepted three-prong tau 
decays $N_{3p}$.
The second term on the right hand side has been added to
take into account 
background carrying no $\tau$ lifetime information.
The background fraction, $f_B$, is estimated from Monte Carlo
and fixed in the fit.
The likelihood
function $B$ is evaluated from
the convolution of a Dirac delta function with the experimental 
resolution function.

The fit minimises $-\log \cal L$.
Average decay
lengths of $\langle L \rangle\,=\,(\,2.245\,\pm\,0.037\,)~\mbox{mm}$ and 
$\langle L \rangle\,=\,(\,2.265\,\pm\,0.051\,)~\mbox{mm}$ are determined for data 
from 1994 and 1995, respectively. 
The results of the fit are represented by the solid lines in 
Figure~\ref{fig:dl}.

The tau lifetime and average decay length are related through the following 
expression:
\begin{equation}
\tau_{\tau}\,=\,\frac{\langle L \rangle}{\beta\,\gamma\,c}.
\end{equation}
Using this equation and taking into account the effects of radiation,
lifetimes of $\tlife\,=\,292.5\,\pm\,4.9~\fs$ 
and $\tlife\,=\,295.2\,\pm\,6.6~\fs$
are obtained for the two data sets,
where the errors are statistical only.

In order to check the decay length method, the analysis is repeated
on a sample of Monte Carlo events. 
The difference between the input tau lifetime and the result of the fit
is assigned as a systematic error due to 
the method.
Systematic effects due to the non-ideal description of the resolution
function are estimated from a one $\sigma$ variation of its parameters.
The deviation of the SMD radius from its nominal
value is measured using
Bhabha and dimuon events by minimising
the track coordinate residuals
from overlapping silicon sensors. Deviations of
$(-5\pm5)~\micron$ and $(-3\pm5)~\micron$
are found in the data of 1994 and 1995, respectively.  
These results are compatible with zero, and
their uncertainties are translated into a systematic error
on the lifetime.
Uncertainties in the fraction of background events carrying no lifetime 
information are estimated from a $\pm\,50\%$ variation
of the fraction. 
Finally, the systematic error due to
the fit range is estimated from the combined data by
a 10\% variation of their
lower and upper bounds. This error also includes the effect of a variation 
of the cut on
$P(\chi^2)$ between 0.1 and 1.5\%. 
Table~\ref{tab:dlsys} summarises the systematic errors
for the decay length method.

The lifetime measurements from the $6620$ three-prong decays from 1994 
and 1995 are combined.
The systematic error due to the resolution function description is 
taken to be uncorrelated; for the other errors a 100\% correlation is
assumed. The result is
$\tlife \,=\,292.9\,\pm\,
                        3.9~\mbox{(stat)}\,\pm\,
                        1.9~\mbox{(syst)}~\fs.
$

\section{Impact Parameter Method}

A second measurement of the tau lifetime is obtained from the 
impact parameter in the plane perpendicular to the beam axis
for one-prong tau decays.
The impact parameter is the distance of closest approach, $\delta$,
of the track to the tau lepton production point, which is estimated by 
the beam position.
In order to be sensitive to the lifetime, a sign is given to the impact
parameter, according to the position of the
intersection between the track and the
tau direction of flight
with respect to the beam position. 
The uncertainty on the impact parameter, $\sip$,
is described
as the quadratic sum of the intrinsic detector resolution, $\sint$, the
size of the interaction region, $(\sxb,\,\syb)$, and a
momentum dependent multiple Coulomb 
scattering contribution, $\sms$,

\begin{equation}
\label{eq:iperror}
\sip^2\,=\,\sint^2\,+\,\sxb^2 \sin^2 \phi\,+\,\syb^2\cos^2 \phi\,+\,
             \smssq,
\end{equation}

\noindent
where $\phi$ is the azimuthal angle in the plane perpendicular
to the beam axis.
The average values for the interaction region size 
are determined from Bhabha and dimuon data and listed in
Table~\ref{tab:bspot}.  

In contrast to the
decay length method a more complicated
function
for the description of the tau decay
is expected here, since
for a one-prong tau decay the decay vertex is {\it a priori} not known. 
From a Monte Carlo study of
the impact parameter distribution at generator level
it is found that
this function can be described in
terms of three exponentials for positive and three
exponentials for negative impact parameter values
\begin{equation}
\label{eq:udl}
U(\delta)\,=\,(1-W)\sum_{i=1}^{3} 
    \frac{f_i^+}{\lambda_i^+} e^{-\frac{\delta}{\lambda_i^+}} \,+\,
                 W \sum_{i=1}^{3}
    \frac{f_i^-}{\lambda_i^-} e^{\frac{\delta}{\lambda_i^-}}.
\end{equation}
In this equation, $W$ represents the fraction of negative impact parameter
values, which originate from an imperfect reconstruction of the tau flight 
direction. 
The slopes of the exponentials, $\lambda_i$,
contain the lifetime
dependence of the distribution. 

\begin{figure*}[hbt]
    \includegraphics[width=0.49\textwidth]{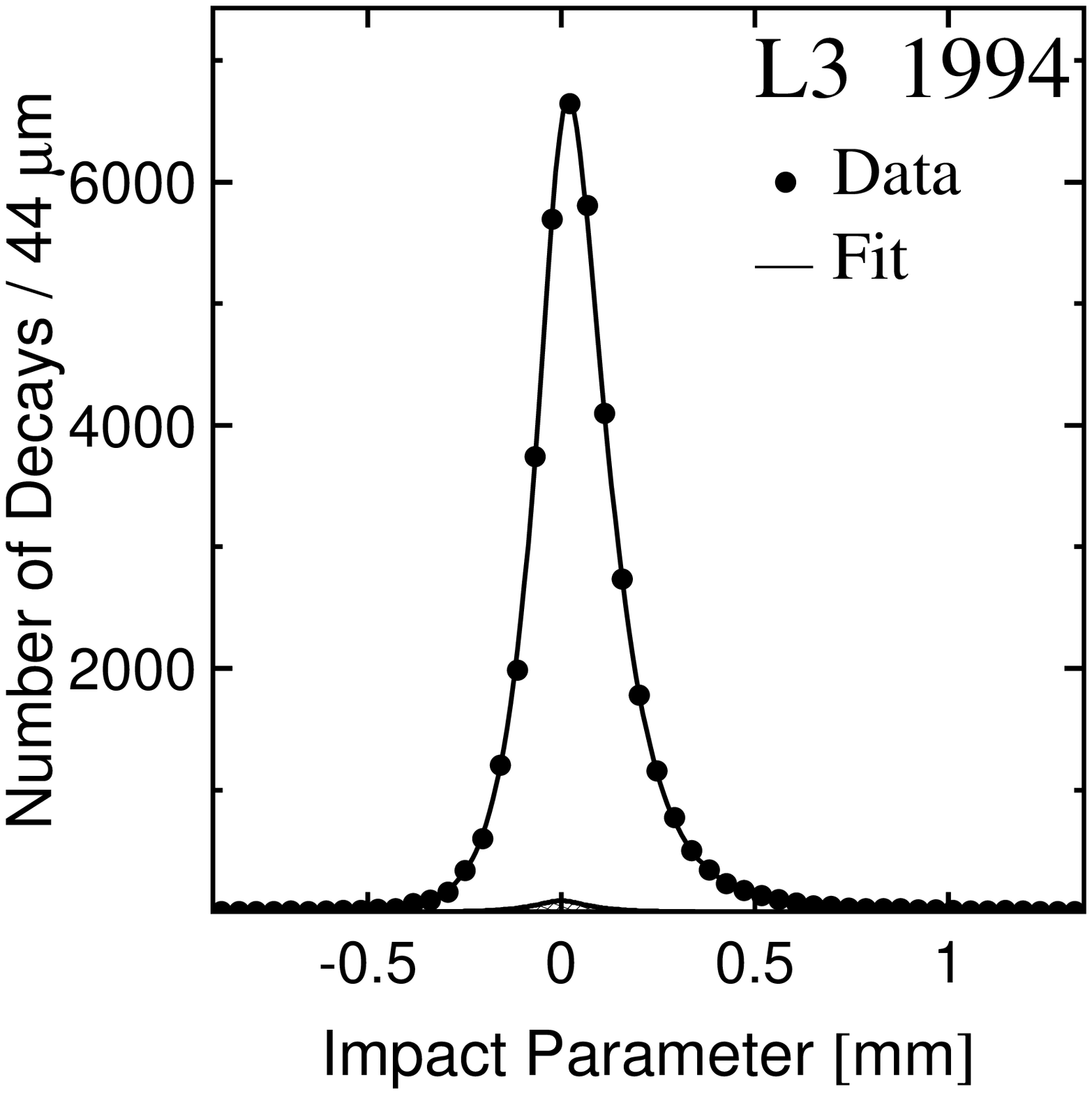}
    \includegraphics[width=0.49\textwidth]{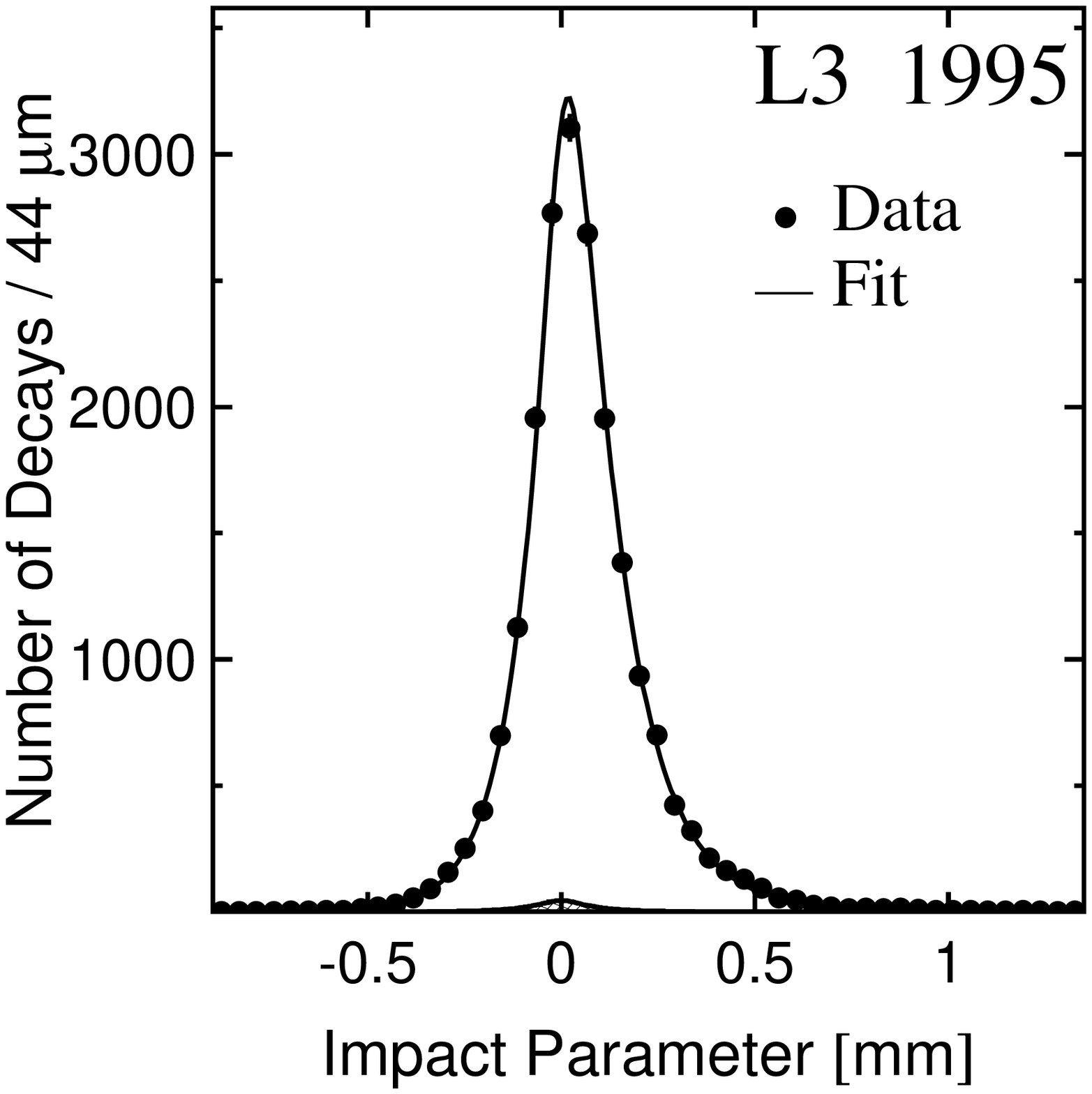}
    \includegraphics[width=0.49\textwidth]{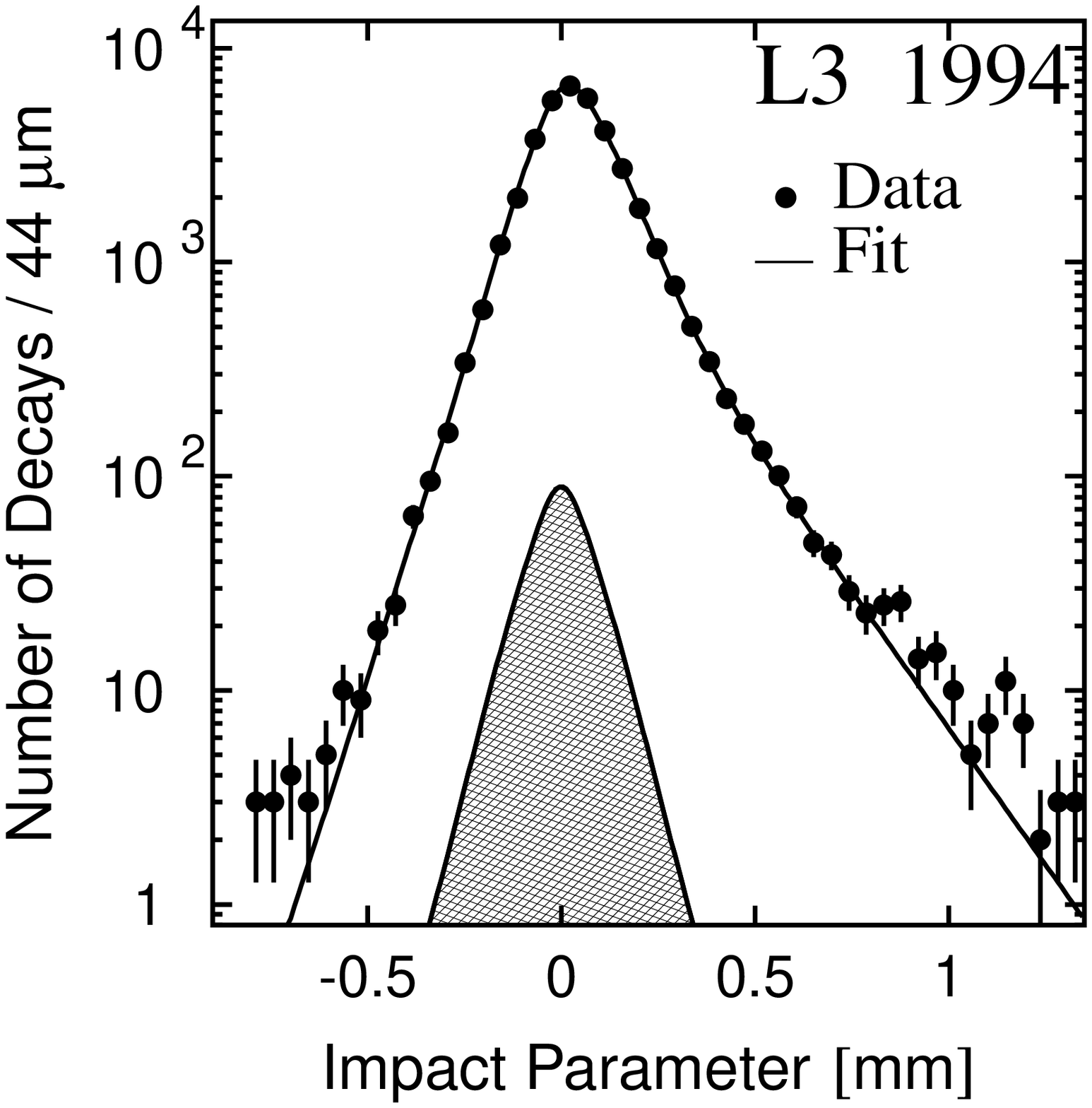}
    \includegraphics[width=0.49\textwidth]{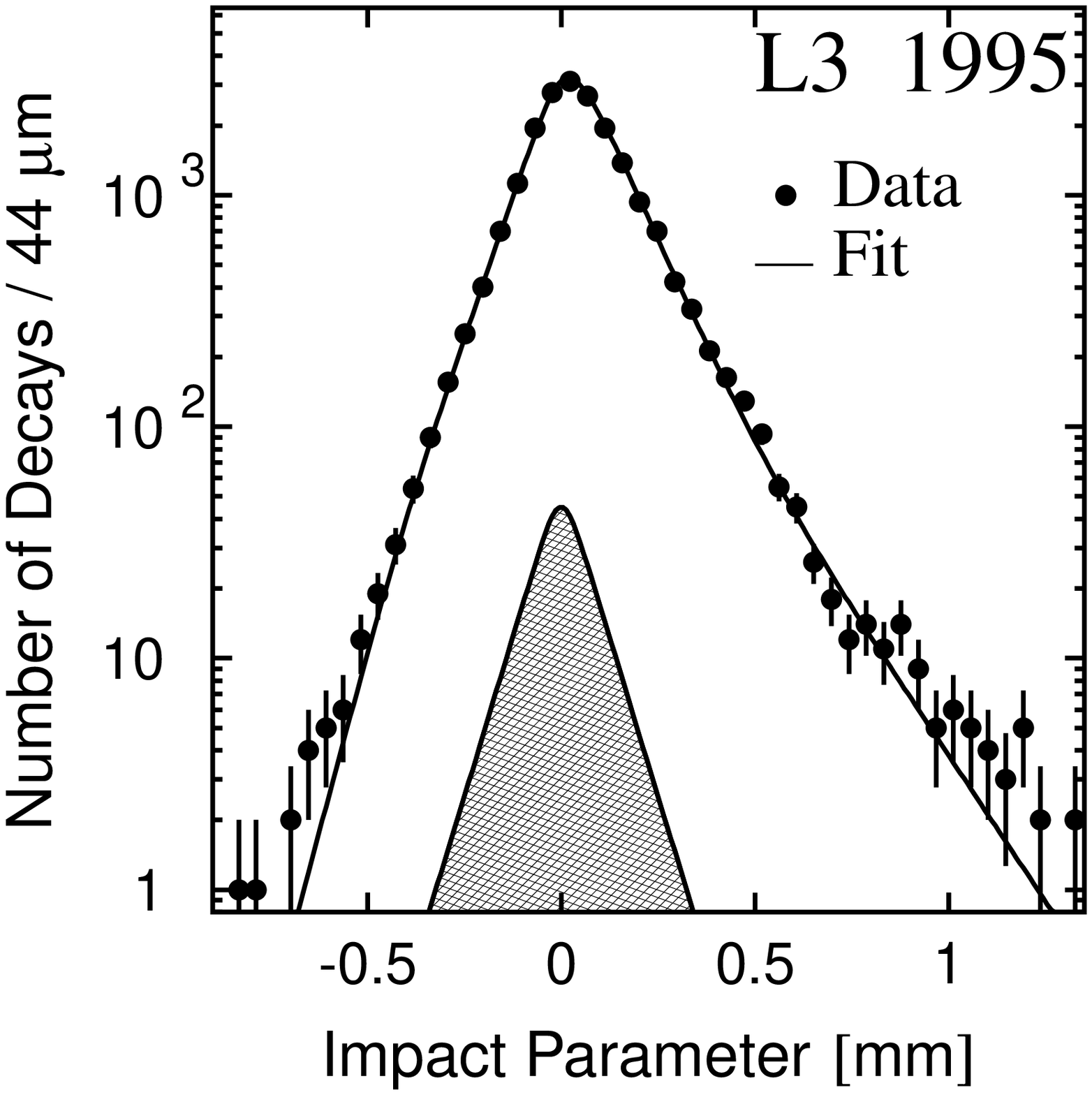}
    \caption[]{\label{fig:ip}
             Impact parameter distributions from 1994~(left) and 
             1995~(right); One-prong tau decays
             are shown in linear scale (upper)
             and logarithmic scale (lower).
             The hatched areas represent the distributions
             of background events which carry no lifetime information.
        }
\end{figure*}

As for the decay length method, the lifetime is extracted from 
an unbinned maximum likelihood fit to the observed distribution.
The likelihood is now determined
from the convolution of a double 
Gaussian resolution function, which is obtained from
Bhabha and dimuon samples, with the function of Eqn.(9). 
The fit also accounts for background carrying no lifetime 
information.
\noindent
Figure~\ref{fig:ip} shows the impact parameter distributions from tau
decays collected in 1994 and 1995. Impact parameters with a value in
the range 
$[-0.9,\,1.35]~\mbox{mm}$ and an impact 
parameter error of
less than $250~\micron$ are accepted for the measurement.
The fit yields a
tau lifetime of $\tlife\,=\,292.7\,\pm\,3.3\,\fs$ and
$\tlife\,=\,295.0\,\pm\,4.9~\fs$ for the two data samples. The errors
are statistical only.

The method is checked on a  Monte Carlo sample, from
which the lifetime is determined in the same way as for the data.
The difference between the input tau lifetime and the result of the fit
is assigned as a systematic error due to 
the method.
Systematic effects due to the uncertainty of the resolution
function are estimated from a variation of its parameters according
to their errors, with correlations taken into account. The beam spot
size is varied according to its statistical errors. The change
in the central value is assigned as a systematic uncertainty. 
The effect of the average SMD radial position 
uncertainty is treated in the same 
way as in the decay length analysis.
The systematic effect due to the knowledge of the function
$U(\delta)$
is evaluated by taking into account its statistical uncertainty and its
dependence on the tau lifetime in the range from 250 to 350~$\fs$. 
The uncertainty arising from the fraction of background 
events is 
estimated from a $\pm\,50\%$ variation of this fraction. 
The systematic error induced by the choice of the fit ranges is
estimated from the combined data sample
by a 10\% variation of their bounds.
Table~\ref{tab:ipsys} summarises the systematic errors
for the impact parameter method. 

The lifetime measurements from the $58656$ one-prong decays from 1994
and 1995 are combined.
The systematic error due to the resolution function is 
taken to be uncorrelated. For the other errors a 100\% correlation has
been assumed. The result is
$\tlife\,=\,292.8\,\pm\,
                        2.7~\mbox{(stat)}\,\pm\,
                        2.0~\mbox{(syst)}~\fs.
$

\section{Discussion}

The combination of the results obtained by the two methods 
with our previous ones~\cite{l3_1993_3,*mblife1} yields
\begin{equation}
\tlife\,=\,293.2\,\pm\,2.0~\mbox{(stat)}\,\pm\,1.5~\mbox{(syst)}~\fs.
\end{equation}
Correlations within the systematic errors are taken into account.
This result supersedes all previous
results~\cite{l3_1993_3,*mblife1,life94}.
This value is in good agreement with the current world
average~\cite{pdg98}.

The measurements of the
branching fractions, $\brel=(17.806~\pm~0.129)\%$ 
and $\brmu=(17.341~\pm~0.129)\%$~\cite{brnew}
together with this lifetime measurement and
the muon lifetime~\cite{pdg98}
yield
$\gwt/\gwe=0.996~\pm~0.006$ and $\gwt/\gwm=0.996~\pm~0.006$
supporting 
the universality hypothesis. 

From the tau lifetime, the tau mass, the muon mass and muon 
lifetime, $\rtau$ is found to be
$\rtau=3.595~\pm~0.048$. This corresponds
to
\begin{equation}
 \als(\mtau^2)=0.319~\pm~0.015~\mbox{(exp)}~\pm~0.014~\mbox{(theory)}.
\end{equation}
The
first error is due to the errors of the 
tau lifetime and the CKM matrix elements~\cite{pdg98}.
The second error is the quadratic sum of
the uncertainties 
resulting from the
renormalisation scale, the term fourth order
in $\als$, the electroweak 
corrections $\sew$, and the non-perturbative
correction, $\dnpqcd$. The renormalisation scale uncertainty is estimated
following Ref.~\cite{pich92} by a
variation
between 0.4 $\leq m_{\tau}^2/\mu^2 \leq$ 2.0 and is the dominant contribution
to the error. 
Other contributions to the theory error
as discussed in Ref.~\cite{anr}
are not considered.
This result is in good agreement with other measurements
of $\als$ at the tau mass~\cite{otherals,pdg98}.
The value of $\als(\mtau^2)$ is extrapolated to the $\Zo$ mass
using the renormalisation group equation~\cite{rge1}
with the four loop calculation of the
QCD $\beta$-functions~\cite{rge}. 
The result, $\alpha_s(\mz^2)$ = 0.1185 $\pm$ 0.0019~\mbox{(exp)}
    $\pm$ 0.0017~\mbox{(theory)}, is in good agreement 
with the current world average value~\cite{pdg98}.

\section*{Acknowledgments}

We thank G. Altarelli and A. Kataev for discussions about the 
estimation of the theoretical uncertainty of $\rtau$.
We wish to
express our gratitude to the CERN accelerator divisions for
the excellent performance of the LEP machine.
We acknowledge the contributions of the engineers
and technicians who have participated in the construction
and maintenance of this experiment.

\newpage

%
%
%
%
%

\newcount\tutecount  \tutecount=0
\def\tutenum#1{\global\advance\tutecount by 1 \xdef#1{\the\tutecount}}
\def\tute#1{$^{#1}$}
\tutenum\aachen            
\tutenum\nikhef            
\tutenum\mich              
\tutenum\lapp              
\tutenum\basel             
\tutenum\lsu               
\tutenum\beijing           
\tutenum\berlin            
\tutenum\bologna           
\tutenum\tata              
\tutenum\ne                
\tutenum\bucharest         
\tutenum\budapest          
\tutenum\mit               
\tutenum\debrecen          
\tutenum\florence          
\tutenum\cern              
\tutenum\wl                
\tutenum\geneva            
\tutenum\hefei             
\tutenum\seft              
\tutenum\lausanne          
\tutenum\lecce             
\tutenum\lyon              
\tutenum\madrid            
\tutenum\milan             
\tutenum\moscow            
\tutenum\naples            
\tutenum\cyprus            
\tutenum\nymegen           
\tutenum\caltech           
\tutenum\perugia           
\tutenum\cmu               
\tutenum\prince            
\tutenum\rome              
\tutenum\peters            
\tutenum\potenza           
\tutenum\salerno           
\tutenum\ucsd              
\tutenum\santiago          
\tutenum\sofia             
\tutenum\korea             
\tutenum\alabama           
\tutenum\utrecht           
\tutenum\purdue            
\tutenum\psinst            
\tutenum\zeuthen           
\tutenum\eth               
\tutenum\hamburg           
\tutenum\taiwan            
\tutenum\tsinghua          
{
\parskip=0pt
\noindent
{\bf The L3 Collaboration:}
\ifx\selectfont\undefined
 \baselineskip=10.8pt
 \baselineskip\baselinestretch\baselineskip
 \normalbaselineskip\baselineskip
 \ixpt
\else
 \fontsize{9}{10.8pt}\selectfont
\fi
\medskip
\tolerance=10000
\hbadness=5000
\raggedright
\hsize=162truemm\hoffset=0mm
\def\r{\rlap,}
\noindent

M.Acciarri\r\tute\milan\
P.Achard\r\tute\geneva\ 
O.Adriani\r\tute{\florence}\ 
M.Aguilar-Benitez\r\tute\madrid\ 
J.Alcaraz\r\tute\madrid\ 
G.Alemanni\r\tute\lausanne\
J.Allaby\r\tute\cern\
A.Aloisio\r\tute\naples\ 
M.G.Alviggi\r\tute\naples\
G.Ambrosi\r\tute\geneva\
% ...
\newpage
\begin{list}{A}{\itemsep=0pt plus 0pt minus 0pt\parsep=0pt plus 0pt minus 0pt
                \topsep=0pt plus 0pt minus 0pt}
\item[\aachen]
 I. Physikalisches Institut, RWTH, D-52056 Aachen, FRG$^{\S}$\\
 III. Physikalisches Institut, RWTH, D-52056 Aachen, FRG$^{\S}$
\item[\nikhef] National Institute for High Energy Physics, NIKHEF, 
     and University of Amsterdam, NL-1009 DB Amsterdam, The Netherlands
\item[\mich] University of Michigan, Ann Arbor, MI 48109, USA
\item[\lapp] Laboratoire d'Annecy-le-Vieux de Physique des Particules, 
     LAPP,IN2P3-CNRS, BP 110, F-74941 Annecy-le-Vieux CEDEX, France
\item[\basel] Institute of Physics, University of Basel, CH-4056 Basel,
     Switzerland
\item[\lsu] Louisiana State University, Baton Rouge, LA 70803, USA
\item[\beijing] Institute of High Energy Physics, IHEP, 
  100039 Beijing, China$^{\triangle}$ 
\item[\berlin] Humboldt University, D-10099 Berlin, FRG$^{\S}$
\item[\bologna] University of Bologna and INFN-Sezione di Bologna, 
     I-40126 Bologna, Italy
\item[\tata] Tata Institute of Fundamental Research, Bombay 400 005, India
\item[\ne] Northeastern University, Boston, MA 02115, USA
\item[\bucharest] Institute of Atomic Physics and University of Bucharest,
     R-76900 Bucharest, Romania
\item[\budapest] Central Research Institute for Physics of the 
     Hungarian Academy of Sciences, H-1525 Budapest 114, Hungary$^{\ddag}$
\item[\mit] Massachusetts Institute of Technology, Cambridge, MA 02139, USA
\item[\debrecen] KLTE-ATOMKI, H-4010 Debrecen, Hungary$^\P$
\item[\florence] INFN Sezione di Firenze and University of Florence, 
     I-50125 Florence, Italy
\item[\cern] European Laboratory for Particle Physics, CERN, 
     CH-1211 Geneva 23, Switzerland
\item[\wl] World Laboratory, FBLJA  Project, CH-1211 Geneva 23, Switzerland
\item[\geneva] University of Geneva, CH-1211 Geneva 4, Switzerland
\item[\hefei] Chinese University of Science and Technology, USTC,
      Hefei, Anhui 230 029, China$^{\triangle}$
\item[\seft] SEFT, Research Institute for High Energy Physics, P.O. Box 9,
      SF-00014 Helsinki, Finland
\item[\lausanne] University of Lausanne, CH-1015 Lausanne, Switzerland
\item[\lecce] INFN-Sezione di Lecce and Universit\'a Degli Studi di Lecce,
     I-73100 Lecce, Italy
\item[\lyon] Institut de Physique Nucl\'eaire de Lyon, 
     IN2P3-CNRS,Universit\'e Claude Bernard, 
     F-69622 Villeurbanne, France
\item[\madrid] Centro de Investigaciones Energ{\'e}ticas, 
     Medioambientales y Tecnolog{\'\i}cas, CIEMAT, E-28040 Madrid,
     Spain${\flat}$ 
\item[\milan] INFN-Sezione di Milano, I-20133 Milan, Italy
\item[\moscow] Institute of Theoretical and Experimental Physics, ITEP, 
     Moscow, Russia
\item[\naples] INFN-Sezione di Napoli and University of Naples, 
     I-80125 Naples, Italy
\item[\cyprus] Department of Natural Sciences, University of Cyprus,
     Nicosia, Cyprus
\item[\nymegen] University of Nijmegen and NIKHEF, 
     NL-6525 ED Nijmegen, The Netherlands
\item[\caltech] California Institute of Technology, Pasadena, CA 91125, USA
\item[\perugia] INFN-Sezione di Perugia and Universit\'a Degli 
     Studi di Perugia, I-06100 Perugia, Italy   
\item[\cmu] Carnegie Mellon University, Pittsburgh, PA 15213, USA
\item[\prince] Princeton University, Princeton, NJ 08544, USA
\item[\rome] INFN-Sezione di Roma and University of Rome, ``La Sapienza",
     I-00185 Rome, Italy
\item[\peters] Nuclear Physics Institute, St. Petersburg, Russia
\item[\potenza] INFN-Sezione di Napoli and University of Potenza, 
     I-85100 Potenza, Italy
\item[\salerno] University and INFN, Salerno, I-84100 Salerno, Italy
\item[\ucsd] University of California, San Diego, CA 92093, USA
\item[\santiago] Dept. de Fisica de Particulas Elementales, Univ. de Santiago,
     E-15706 Santiago de Compostela, Spain
\item[\sofia] Bulgarian Academy of Sciences, Central Lab.~of 
     Mechatronics and Instrumentation, BU-1113 Sofia, Bulgaria
\item[\korea]  Laboratory of High Energy Physics, 
     Kyungpook National University, 702-701 Taegu, Republic of Korea
\item[\alabama] University of Alabama, Tuscaloosa, AL 35486, USA
\item[\utrecht] Utrecht University and NIKHEF, NL-3584 CB Utrecht, 
     The Netherlands
\item[\purdue] Purdue University, West Lafayette, IN 47907, USA
\item[\psinst] Paul Scherrer Institut, PSI, CH-5232 Villigen, Switzerland
\item[\zeuthen] DESY, D-15738 Zeuthen, 
     FRG
\item[\eth] Eidgen\"ossische Technische Hochschule, ETH Z\"urich,
     CH-8093 Z\"urich, Switzerland
\item[\hamburg] University of Hamburg, D-22761 Hamburg, FRG
\item[\taiwan] National Central University, Chung-Li, Taiwan, China
\item[\tsinghua] Department of Physics, National Tsing Hua University,
      Taiwan, China
\item[\S]  Supported by the German Bundesministerium 
        f\"ur Bildung, Wissenschaft, Forschung und Technologie
\item[\ddag] Supported by the Hungarian OTKA fund under contract
numbers T019181, F023259 and T024011.
\item[\P] Also supported by the Hungarian OTKA fund under contract
  numbers T22238 and T026178.
\item[$\flat$] Supported also by the Comisi\'on Interministerial de Ciencia y 
        Tecnolog{\'\i}a.
\item[$\sharp$] Also supported by CONICET and Universidad Nacional de La Plata,
        CC 67, 1900 La Plata, Argentina.
\item[$\diamondsuit$] Also supported by Panjab University, Chandigarh-160014, 
        India.
\item[$\triangle$] Supported by the National Natural Science
  Foundation of China.
\item[\dag] Deceased.
\end{list}
}
\vfill


\newpage
\begin{table}[tbh]
  \begin{center}
    \begin{tabular}{|l|c|c|}
      \hline     
      Error Source~$(\fs)$  & 1994    & 1995    \\
      \hline
      Method                & 0.5     & 0.5     \\
      Resolution function   & 1.5     & 2.0     \\ 
      SMD radius            & 0.7     & 0.7     \\
      Background estimate   & 0.8     & 0.6     \\
      Fit range             & \multicolumn{2}{c|}{1.0}   \\
      \hline
     \end{tabular}
    \caption{Systematic errors for the decay length method}
    \label{tab:dlsys}
  \end{center}
\end{table}

\begin{table}[tbh]
  \begin{center}
    \begin{tabular}{|l|c|c|}
      \hline
      &  $\sxb~(\micron)$          &  $\syb~(\micron)$ \\
      \hline
      1994        &  $118\,\pm\,1$             &  $14\,\pm\,1$     \\
      1995        &  $148\,\pm\,2$             &  $15\,\pm\,3$     \\
      \hline
    \end{tabular}
    \caption{The size of the interaction regions}
    \label{tab:bspot}
  \end{center}
\end{table}
\begin{table}[tbh]
  \begin{center}
  \begin{tabular}{|l|c|c|}
    \hline     
Error Source~$(\fs)$  & 1994    & 1995    \\
    \hline
Method                & 0.2     & 0.2     \\
Resolution function   & 1.1     & 1.5     \\  
Beam spot size        & 0.5     & 0.5     \\
SMD radius            & 0.7     & 0.7     \\
Function $U(\delta)$  & 1.3     & 1.3     \\
Background estimate   & 0.5     & 0.5     \\
Fit range             & \multicolumn{2}{c|}{0.5}     \\
 \hline
  \end{tabular}
  \caption{Systematic errors for the impact parameter method}
  \label{tab:ipsys}
  \end{center}
\end{table}

\newpage

\end{document}